\documentclass[aps,pra,floatfix,twocolumn,superscriptaddress,nofootinbib,showpacs]{revtex4} 

\usepackage{graphicx}
\usepackage{graphics}
\usepackage{amssymb}
\usepackage{amsmath}
\usepackage{epsfig}
\usepackage{latexsym}
\usepackage{color}

\newcommand{\bra}[1]{\left\langle{#1}\right\vert}
\newcommand{\ket}[1]{\left\vert{#1}\right\rangle}

\newcommand{\beq}{\begin{equation}}
\newcommand{\eeq}{\end{equation}}
\newcommand{\bqa}{\begin{eqnarray}}
\newcommand{\eqa}{\end{eqnarray}}
\newcommand{\nn}{\nonumber}

\newcommand{\erf}[1]{Eq.~(\ref{#1})}
\newcommand{\dg}{^\dagger}

\begin{document}

\title{Continuous quantum error correction by cooling}

\author{Mohan Sarovar}
\email{mohan@physics.uq.edu.au} \affiliation{Centre for Quantum
Computer Technology, and School of Physical Sciences, The
University of Queensland, St Lucia, QLD 4072, Australia}

\author{G. J. Milburn}
\email{milburn@physics.uq.edu.au} \affiliation{Centre for Quantum
Computer Technology, and School of Physical Sciences, The
University of Queensland, St Lucia, QLD 4072, Australia}


\begin{abstract}
We describe an implementation of quantum error correction that operates continuously in time and requires no active interventions such as measurements or gates. The mechanism for carrying away the entropy introduced by errors is a cooling procedure. We evaluate the effectiveness of the scheme by simulation, and remark on its connections to some recently proposed error prevention procedures.
\end{abstract}

\maketitle


\section{Introduction}
Error correction and prevention will most likely have a major role to play in the operation of any future quantum information processing or storage device. Since the discovery of quantum error correcting codes (ECCs) by Shor \cite{Shor_1995} and Steane \cite{Steane_1996}, there has been much activity on the development of new error correction and prevention techniques. These techniques can be broadly split into two types: the passive schemes that exploit dynamical symmetries to encode quantum information in \textit{noiseless subsystems} \cite{Lidar_W_2003}, and the active schemes that involve the continued execution of operations to suppress the buildup of errors. The active schemes can be further split into two subclasses: \textit{open-loop}, error prevention schemes (e.g. dynamic decoupling, bang-bang control) that are based on controlling the interaction between the system and the error inducing environment \cite{Viola_KL_1999, Zanardi_1999, Duan_G_1999, Facchi_TPNTL_2004, Viola_2004}, and \textit{closed-loop}, error correction schemes that use ECCs. We shall be concerned with the active, closed-loop, error correction techniques in this paper.

There are two ways to implement such active error correction schemes that use ECCs - with and without measurement \cite{mikeandike} - and standard prescriptions for implementing both alternatives require ideal resources such as projective measurements, instantaneous unitary gates, and fast resetting operations. What if these resources are not available? For many current quantum computing architectures, some subset of these ideal operations will not be available in the near future. So the question we address is: can one effectively perform error correction without these ideal operations? This question was examined for the case of active error correction schemes that use measurement in \cite{Ahn_DL_2002, Sarovar_AJM_2004}, and in this paper we will concentrate on the other case: active error correction \textit{without} measurement. 

We replace the instantaneous gates and reset operations necessary for error correction without measurement (ECWM) with more modest resources and apply them in a continuous manner. This results in a scheme for error correction which is \textit{automatic} in the sense that no external actions are needed, and has a description in terms of continuous time dynamical maps. An example of such a dynamical map is solved numerically to evaluate the effectiveness of such implementations. We discuss the scheme primarily in the context of quantum memory where the preservation of quantum information is the aim rather than computation. The implementation is most applicable in this context because of its automatic and continuous nature. We do not consider coded logical operations during the error correction process.

The paper is organized as follows: section \ref{sec:ec_wo_meas} introduces error correction without measurement and presents an example that we shall use in the remainder of the paper. Section \ref{sec:ec_cont} transforms this description into a continuous version that uses non-ideal resources and presents an analysis of its performance. Section \ref{sec:impl} examines possibilities of physically implementing the scheme, and we conclude with a discussion in section \ref{sec:disc}.

\section{Error correction without measurement}
\label{sec:ec_wo_meas}

\subsection{The error model}
\label{sec:errmodel}
Before describing particular error correction schemes it is important to outline the exact error model being treated. We consider a scenario where unitary error operators act at randomly distributed times and independently on each qubit of the encoded state. In addition, the probability of an error is independent of the state of the system. This is a fairly standard error model in the error correction literature \cite{mikeandike} and is realistic if the major source of noise is coupling to a large Markovian environment. 

A continuous time description of a system under such an error model is the following master equation for the dynamics of the system density operator
\begin{equation}
\label{eq:err_model}
\frac{d\rho}{dt} = \sum_i \gamma_i\mathcal{D}[U_i]\rho
\end{equation}
where $U_i$ are the unitary error operators and $\mathcal{D}$ is the superoperator
\begin{equation}
\mathcal{D}[A]\rho = A\rho A\dg - \frac{1}{2}A\dg A\rho - \frac{1}{2}\rho A\dg A 
\end{equation}
for any operator $A$. $\gamma_i$ are the rates for each of the error operators. That is, the average number of errors of type $i$ in a time $dt$ is $\gamma_idt$.

\subsection{Error correction using codes}
Closed-loop error correction schemes use error correction codes to introduce redundancy in such a manner that in a certain subspace - the \textit{codespace} - of the total system Hilbert space a certain subset of errors become reversible. The procedure for reversing these errors typically involves a detection step that calculates whether or not an error occurred (referred to as calculating the \textit{error syndrome}), followed by a correction step that reverses its effect. In implementations of error correction that do not use measurement, these two steps are done by coupling the encoded system to ancilla qubits. This coupling performs the detection step by putting the value of the error syndrome in these ancilla qubits, and the correction step by conditionally applying gates to the encoded system, conditioned on the ancilla qubit values. 

We will illustrate this process by using a simple example that implements a code to protect against \textit{bit-flip} errors. A bit-flip error reverses the value of qubit computational basis states - i.e. $\ket{0} \rightarrow \ket{1}$ and $\ket{1} \rightarrow \ket{0}$ under the action of the error. The \textit{bit-flip code}, which is an example of a wide class of codes called \textit{stabilizer codes} \cite{Gottesman_thesis, mikeandike}, protects against this error by using the following repetition encoding: $\ket{0}_L \equiv \ket{000}_P, ~ \ket{1}_L \equiv \ket{111}_P$, where the subscripts L and P stand for logical and physical, respectively. Therefore a general encoded qubit will have the form $\ket{\psi} = \alpha\ket{0}_L + \beta\ket{1}_L = \alpha\ket{000}_P + \beta\ket{111}_P$ with $|\alpha|^2 + |\beta|^2 = 1$. The encoded qubit states are referred to as the \textit{codewords}, and the subspace they span as the \textit{codespace}.

This code can detect and correct one bit-flip. The detection operation involves measuring the operators $ZZI$ and $IZZ$ \footnote{We denote the Pauli $\sigma_X, \sigma_Y, \textrm{and } \sigma_Z$ operators by $X$, $Y$, and $Z$, respectively, and suppress the tensor product sign. Therefore $ZZI \equiv \sigma_Z \otimes \sigma_Z \otimes I$.}, which are referred to as the \textit{error syndromes}. Two things to note, both of which are properties of all stabilizer codes, are that all the error syndrome operators commute with each other, and that the codewords are both eigenvalue one eigenstates of the syndromes (or in other words, the codespace is stabilized by the syndrome operators).

The four possible outcomes of the two syndrome measurements label the four possible error events. This is illustrated by table \ref{tab:tq_table}. Correcting errors using this code then simply amounts to applying a unitary to restore the encoded state back to its unperturbed value. The value of this unitary depends on the measurement results as table \ref{tab:tq_table} shows.

\begin{table}
\begin{tabular}{|c|c|c|c|}
  \hline
  \textbf{$\langle ZZI\rangle_\rho$} & \textbf{$\langle IZZ\rangle_\rho$} & \textbf{Error} & \textbf{Correcting unitary} \\
  \hline\hline
  +1 & +1 & None & None \\
  -1 & +1 & on qubit 1 & XII \\
  +1 & -1 & on qubit 3 & IIX \\
  -1 & -1 & on qubit 2 & IXI \\
  \hline
\end{tabular}
\caption{The three qubit bit-flip code. Note that each error results in a different sequence of error syndromes. $\langle \cdot \rangle_\rho$ represents the expectation value of $\cdot$ under the encoded three qubit state $\rho$.} \label{tab:tq_table}
\end{table}

\begin{figure}
\includegraphics[scale=0.3]{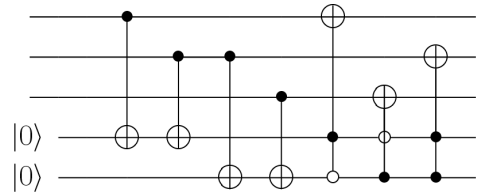}
\caption{A circuit for implementing the three qubit bit-flip code without measurement. The top three qubits form the encoded logical qubit and the bottom two are ancilla. Note that to repeat the error correction procedure, the ancilla qubits must be replaced or reset to the $\ket{0}$ state at the end of each run (at the far right of the circuit).} \label{fig:bf_circuit}
\end{figure}

A circuit that implements this error correction code, and does so without using measurement is given in figure \ref{fig:bf_circuit}. In this circuit, the first three CNOT gates have the effect of calculating the error syndrome operator values (under the encoded state in the top three qubits) and placing them into the ancilla qubits. Then the correction is done by direct coupling between the ancilla and the encoded qubits (via Toffoli gates which provide the ability to condition upon the values of both ancilla qubits). It is important to note that the ancilla qubits must be reset to the $\ket{0}$ state after each run of the circuit. This is a consequence of the fact that the entropy generated by the errors is moved into the ancilla subsystem and must be carried away before the next run of the circuit.

%

This circuit illustrates the essential ideas behind implementing ECWM: introduction of ancilla qubits, their direct coupling to the encoded qubits, and the resetting of these ancilla qubits after each cycle. If this cycle, comprised of detect, correct, and reset is performed often enough, and the only errors in our system are independent bit-flip errors at randomly distributed times, then one can preserve the value of logical qubit indefinitely. Here, `often enough' can be precisely defined as: the interval between subsequent cycles must be small enough so that the probability of two or more bit-flip errors occurring is negligible. If we need to handle a larger set of errors, we would use a more complex code, but the implementation of the error correction would procede in the same manner as in this simple example.

Note that we are assuming that the operations involved in the circuit - the unitary gates and the ancilla reset - are ideal and instantaneous. More precisely, we are assuming that operations take a negligible amount of time with respect to the time scale set by the rate of the bit-flip errors. This is exactly the assumption that we will remove in the next section when we replace these operations by non-instantaneous versions and describe the whole process in a continuous manner. 

Error correction without measurement has interesting connections to the quantum Zeno effect . The whole error correction procedure can be viewed as a method of constraining the dynamics of a system to a two dimensional subspace of Hilbert space by strongly interacting it with a heavily damped ancillary system whose dynamics are Zeno inhibited in the limit of infinite damping. This equivalence between standard implementations of ECWM and the Zeno effect is explored in the appendix of this paper.

\section{The continuous time implementation}
\label{sec:ec_cont}
There are two principal differences between our continuous time implementation of ECWM and the standard discrete model of the last section:
\begin{enumerate}
\item The unitary gates which form the system-ancilla coupling are replaced with an equivalent effective Hamiltonian with finite strength. This Hamiltonian performs both the detection and correction operations continuously and simultaneously.
\item The ancilla reset procedure is replaced with the analogous continuous process of \textit{cooling}. Each ancilla qubit must be independently and continuously cooled to its ground state ($\ket{0}$). Note that this assumes that the fiducial state of the ancilla qubit is the ground state: $\ket{0}$. This is not a restrictive condition because the error correction code can always be modified so that this is the case.  
\end{enumerate}
These changes lead to a continuous time description of the ECWM process in terms of a master equation. This master equation is Markovian because both the open system components - the errors and the ancilla cooling - are Markovian processes.

We illustrate this continuous time implementation by modeling its dynamics for the bit-flip code outlined in the last section. The continuous time description of the circuit of figure \ref{fig:bf_circuit} is:
\begin{eqnarray}
\label{eq:bf_mastereqn}
\frac{d\rho}{dt} &=& \gamma(\mathcal{D}[XIIII] + \mathcal{D}[IXIII] + \mathcal{D}[IIXII])\rho \nn \\
&& + \lambda(\mathcal{D}[IIIS^-I] + \mathcal{D}[IIIIS^-])\rho \nn \\
&& - i\kappa[H, \rho]
\end{eqnarray}
where $\gamma$ is the bit-flip error rate, $\kappa$ is the strength of $H$, the Hamiltonian which performs the detection and correction, and $\lambda$ is the rate of the cooling applied to the ancilla qubits. $S^- \equiv \frac{1}{2}(X+iY) = \ket{0}\bra{1}$ is the qubit lowering operator, and the ordering of the tensor product for all operators in the equation runs down the circuit (i.e. the first three operators apply to the encoded qubit, and the last two to the ancilla). Note that we set $\hbar=1$ throughout the paper. A master equation describing the continuous time implementation of a general code will follow the same pattern: independent cooling for each ancilla required, a Hamiltonian that couples the encoded and ancilla qubits, and decoherence terms for each error of concern.

The Hamiltonian in \erf{eq:bf_mastereqn} is the effective Hamiltonian for the whole unitary gate sequence of figure \ref{fig:bf_circuit}. It can be written explicitly as $H = H_D + H_C + i[H_D, H_C]$ where $H_D$ and $H_C$ are Hamiltonians that perform the detection and correction operations, respectively. That is, only the first term in the Cambell-Baker-Hausdorff expansion \cite{Sternberg_2004} is needed for a good approximation. The explicit forms of $H_D$ and $H_C$ are:

\begin{widetext}
\begin{eqnarray}
\label{eq:ham_brakets}
H_D &=& \ket{00101}\bra{00100} + \ket{11001}\bra{11000} + \ket{ 10010}\bra{10000} \nn \\ &&+ \ket{01110}\bra{01100} + \ket{01011}\bra{01000} + \ket{10111}\bra{10100} + h.c.\nn \\
H_C &=& \ket{00001}\bra{00101} + \ket{11101}\bra{11001} + \ket{00010}\bra{10010} \nn \\ &&+ \ket{11110}\bra{01110} + \ket{00011}\bra{01011} + \ket{11111}\bra{10111} + h.c.
\end{eqnarray}
\end{widetext}

Each term in $H_D$ represents the detection of an error and each term in $H_C$ represents the correction of an error. The Hamiltonian necessary for a general error correction code will follow the same prescription, with appropriate $H_D$ and $H_C$.

Note that in \erf{eq:bf_mastereqn} the error processes are only modeled on qubits that form the encoded state. We can extend the errors dynamics onto the ancilla qubits as well, however, in the parameter regime we shall be interested in - the parameter regime where the error correction is effective - the cooling will dominate all other ancilla dynamics. That is, we shall see that $\lambda \gg \gamma$, and thus we can ignore the error dynamics on the ancilla qubits.

We use this particular example to evaluate the efficacy of this implementation of error correction. We solve \erf{eq:bf_mastereqn} by numerical integration and monitor the evolution of the average fidelity, a figure of merit capturing how well the logical qubit is preserved. The fidelity measure used is simply the overlap with the state to be preserved: $F(t) \equiv \bra{\psi}\rho(t)\ket{\psi}$, where $\rho(t)$ is the reduced state of just the encoded subsystem.

\begin{figure}[]
\includegraphics[scale=0.5]{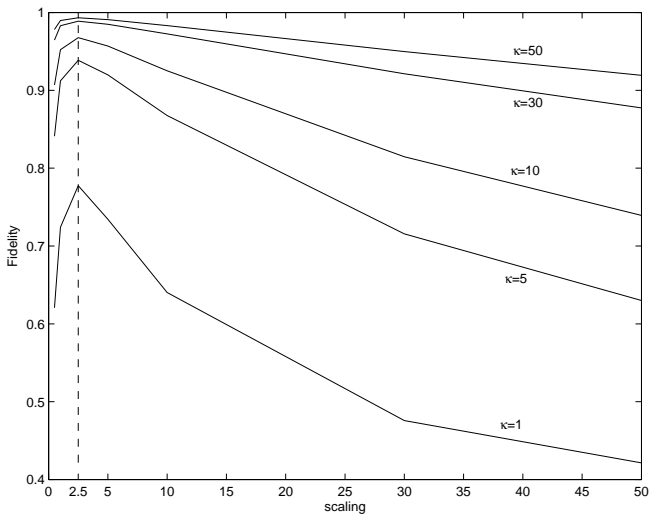}
\caption{Average fidelity, after a fixed period of time (T=10), of an encoded qubit (three qubit code) undergoing continuous error correction. The different curves are for different Hamiltonian strengths ($\kappa$) and the horizontal axis shows how the cooling rate is scaled with $\kappa$; i.e. $\lambda = s\kappa$ where $s$ is varied along the horizontal axis. The units of time are arbitrary. Other parameters: $\gamma=0.05~Hz$, and initial state $\ket{\psi_0} = \ket{000}$. } 
\label{fig:scaling}
\end{figure}

Note that there are three parameters to choose in \erf{eq:bf_mastereqn}: the error rate ($\gamma$), Hamiltonian strength ($\kappa$), and the cooling rate ($\lambda$). We expect the last two to be intimately linked because while $\kappa$ determines the rate at which information is exchanged between the encoded qubits and ancilla qubits, $\lambda$ determines the rate at which this information is carried away from the system. We need a good match between the two if the error correction procedure is to work. From a control systems perspective this is analogous to tuning the parameters of an autonomous controller (e.g. PID controller) to achieve a desired control objective. Figure \ref{fig:scaling} shows the average fidelity after a fixed period of time for several combinations of $\kappa$ and $\lambda$ values and it is clear that the best performance is when $\lambda \approx 2.5\kappa$ \footnote{This optimal point is independent of the error rate and the initial state of the encoded qubits.}. We assume this optimal operating point from here on, reducing the number of free parameters to two.

\begin{figure}[]
\includegraphics[scale=0.5]{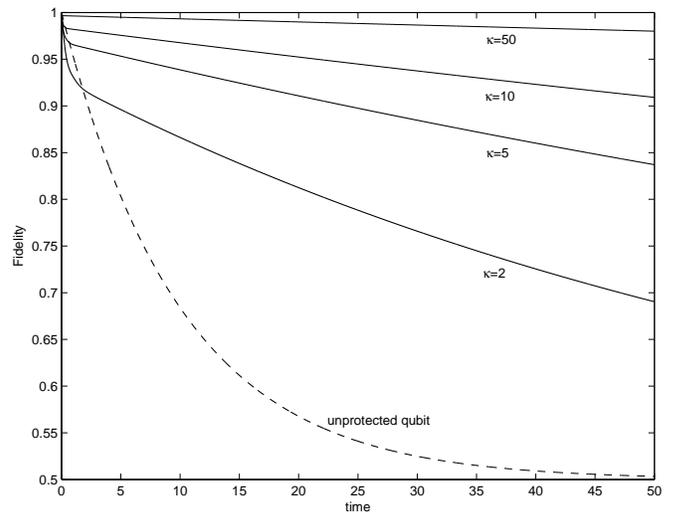}
\caption{Fidelity curves for several Hamiltonian strengths. The solid curves are the average fidelity of an encoded qubit (three qubit code) with continuous error correction (parameters used: $\gamma=0.05~Hz, \lambda=2.5\kappa$, initial state $\ket{\psi_0} = \ket{000}$). The units of time are arbitrary. The dashed curve is the fidelity of one qubit undergoing random bit-flips without error correction (initial state $\ket{\psi_0} = \ket{0}$).} 
\label{fig:fids}
\end{figure}

Figure \ref{fig:fids} shows the evolution of fidelity with time for a fixed error rate and several values of $\kappa$ (with $\lambda$ kept at $2.5\kappa$). This clearly shows an improvement in performance with an increase in the Hamiltonian strength. This agrees with intuition because in the limit of very large $\kappa$, this implementation is the same as the corresponding discrete implementation with the detect-correct-reset cycle operating at a very high frequency.

\begin{figure}[]
\includegraphics[scale=0.5]{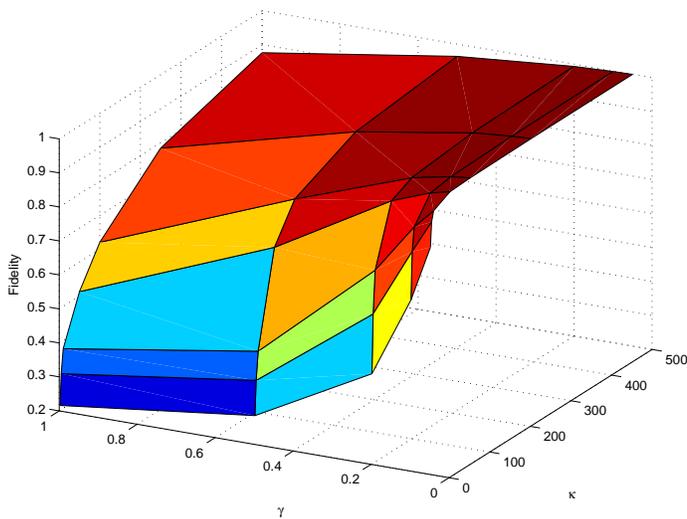}
\caption{(Color online) Average fidelity curves for several combinations of error rate and Hamiltonian strength (parameters used: $\lambda=2.5\kappa$, initial state $\ket{\psi_0} = \ket{000}$).} 
\label{fig:gk_surf}
\end{figure}

We can also characterize the scheme by varying both free parameters ($\gamma$ and $\kappa$) and examining the average fidelity of an encoded state after a fixed period of time. This leads to the surface shown in figure \ref{fig:gk_surf}. As expected, the scheme's performance improves for large values of $\kappa$ and deteriorates for large values of $\gamma$. The figure also suggests that the performance of the scheme does not scale in the same manner with the two parameters. Small increases in $\gamma$ require much larger increases in $\kappa$ (and consequently $\lambda$) to maintain average fidelity values. For example, the fidelity at the point ($\gamma=0.2, \kappa=100$) is poorer than at the point where both parameters are quadrupled: ($\gamma=0.8, \kappa=400$). In effect, the ratio $\gamma/\kappa$ is not sufficient to completely characterize the performance of the scheme. 

\begin{figure}[]
\includegraphics[scale=0.5]{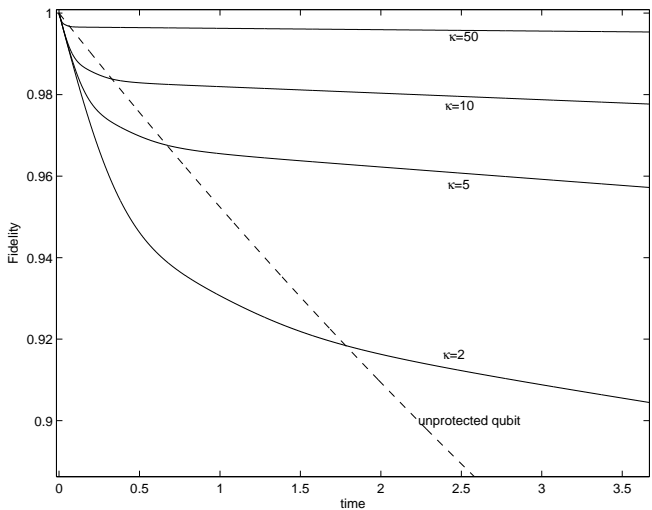}
\caption{Zoomed in version of figure \ref{fig:fids}. Units of time are arbitrary, and the parameter values are the same as for figure \ref{fig:fids}} 
\label{fig:fids_zoom}
\end{figure}

Another interesting aspect of figure \ref{fig:fids} is the behaviour of the fidelity curves shortly after the initial time. A zoomed in version of the figure is shown in figure \ref{fig:fids_zoom}, and it shows that the error corrected system initially performs worse than the uncorrected qubit. In fact, it is during this initial period that the major loss of fidelity occurs; after it the average fidelity decays almost linearly with time. This initial poor performance is because the finite strength Hamiltonian requires some time to recognize and respond to the error process. We can make this fidelity loss arbitrarily small, but at the price of increasing the strength of the Hamiltonian. From a dynamical systems perspective, the amount of fidelity loss is directly related to the amount of delay in the control system, and this decreases with increasing $\kappa$.

We expect analogous continuous time implementations of other codes to exhibit all of the features highlighted above in this bit-flip code example: an optimal operating point in the $\kappa$-$\lambda$ parameter plane, improving performance with increasing Hamiltonian strength and cooling rate, and poor initial time behaviour due to the Hamiltonian nature of the control system. 

\section{Physical Implementation}
\label{sec:impl}
The primary difficulty in implementing such a continuous scheme for error correction lies in manufacturing the Hamiltoninan necessary for the detect and correct operations (e.g. $H$ in \erf{eq:bf_mastereqn} for the bit-flip code). The remaining step in the error correction scheme, the cooling, is an easy and natural operation to implement in most systems.

The Pauli decomposition of the detect and correct Hamiltonians for the bit-flip code (\erf{eq:ham_brakets}) are:

\begin{widetext}
\begin{eqnarray}
\label{eq:ham_pauli}
H_D &=& III(IX + XI + XX + XZ - YY + ZX) + IZZ(-IX + XI - XX + XZ + YY - ZX) \nn \\
&+& ZIZ(-IX - XI + XX - XZ - YY -ZX) + ZZI(IX - XI - XX - XZ + YY + ZX) \nn \\
H_C &=& (IIX + IXI + XII + XZZ + ZXZ + ZZX)II + (-IIX - IXI + XII + XZZ - ZXZ - ZZX)IZ \nn \\
&+& (IIX - IXI - XII - XZZ - ZXZ + ZZX)ZI \nn \\
&+& (-IIX + IXI - XII - XZZ + ZXZ - ZZX)ZZ
\end{eqnarray}
\end{widetext}

Note that when the encoded qubits are in the codespace and the ancilla qubits are in the $\ket{00}$ state, $H_D$ and $H_C$ evaluate to zero as we would expect. From this decomposition, we can see that even for the simple bit flip code, the coupling between the encoded and ancilla qubits is complex and involves many-body terms - such a Hamiltonian would be difficult to manufacture on any of the current quantum computing architectures. 

This error correction scheme is a complex implementation of \textit{coherent quantum feedback}  \cite{Lloyd_2000} with an added dissipative channel. Such systems were also considered by Barnes and Warren \cite{Barnes_W_2000}. This paper can be viewed as relating their scheme, stated in terms of energy principles, to more standard implementations of error correcting codes. And as Barnes and Warren did in \cite{Barnes_W_2000}, we conclude that this new implementation demonstrates that given the ability to manufacture a complex coupling between an encoded system and an ancilliary system, it is possible to perform error correction by a cooling (dissipative) process alone.

With regards to physical implementations, we also note that Beige \textit{et. al.} have proposed error prevention schemes \cite{Beige_BTK_2000, Beige_2003, Beige_2004} for atom-cavity and ion trap architectures that rely on cooling and have similarities to the continuous scheme we detail. Because these proposals perform error prevention as opposed to error correction, the coupling they require between the encoded and ancilla systems is simpler than the ones we require. This makes their schemes implementable on current quantum computing architectures. \footnote{The Beige \textit{et. al} schemes have the additional feature that the dissipation channel enables simplified encoded operations.}

A possible avenue for future work is to investigate simpler encoded-ancilla system couplings that allow error \textit{correction}. If such couplings exist, it would make the physical implementation of error correction through continuos coherent feedback (and cooling) much more feasible.

\section{Discussion}
\label{sec:disc}
We have shown by example that a continuous time implementation of error correction without measurement is effective for preserving quantum information. Due to the continuous and automatic nature of the correcting operation, such implementations are ideal for preserving quantum memory but less suited to error correction during quantum computation. It should be noted that even though we demonstrated the scheme for the three-qubit bit-flip code, it can in principle be used to implement any quantum error correction code.
%


Aside from describing a different implementation of error correction, the scheme above casts error correction in terms of the very natural process of cooling; it refines the viewpoint that error correction is a `cooling process' which extracts the entropy that enters the system through errors. However error correction is not cooling to a particular state such as a ground state, but rather a subspace of Hilbert space, and the specially designed coupling Hamiltonian allows us to implement this cooling to a (non-trivial) subspace by a simple cooling of the ancilla qubits to their ground state.

Finally, a compelling reason to consider such continuous time implementations of error correction is that they give one an idea of how effective error correction can be in the absence of ideal resources. This has been a contentious issue recently \cite{Alicki_2004}, and is of much practical importance. The continuous time implementation sketched in this paper and its counterparts in \cite{Ahn_DL_2002, Sarovar_AJM_2004} provide an upper bound to the performance of error correction schemes that do not have access to instantaneous gates, measurements and reset operations. They provide a method for answering the question: given a certain intrinsic error rate, how fast do the measurements, gates, or reset operations have to be to achieve a desired fidelity criterion?

\section{Acknowledgements}
\label{sec:acks} We gratefully acknowledge the support of the Australian Research Council Centre of Excellence in Quantum Computer Technology. MS would like to thank Michael Nielsen, Michael Bremner, and Charles Hill for helpful discussions.

\appendix
\section{Error correction without measurement and the Zeno effect}
\label{sec:appendix}
It is instructive to recast the standard error correction without measurement implementation of section \ref{sec:ec_wo_meas} in terms of another well known process: the quantum Zeno effect (QZE).  The QZE occurs when the irreversible interaction between the system and a measuring device is so strong that the evolution of the system is confined to a specific subspace \cite{Misra_S_1977, Facchi_P_2001, Facchi_LP_2003}.  The effect of the interaction is to suppress coherence between any  state in the relevant subspace and  states outside the subspace to such a degree that the dynamics can never leave the subspace. For example, if repeated projective measurements of the projector onto the initial state of a dynamical system are made, the probability for the system to not leave the initial state remains arbitrarily close to unity for very long times.  ECWM is precisely this: the resetting of the ancilla qubits (together with their very specific coupling to the encoded qubits) results in a confinement of the encoded state's evolution to the codespace. 

In the measurement version of the Zeno effect, frequent and arbitrarily accurate measurements are modeled by the application of a projection operator onto the subspace, $P$, at periodic intervals to yield a discrete dynamics of the form
\begin{equation}
\ket{\psi (t)} = (Pe^{-iH\frac{t}{N}})^N\ket{\psi_0}
\end{equation}
where $\ket{\psi(t)}$ is the state at time $t$ during which there have been $N$ projections, $H$ is the natural evolution of the system, and $\ket{\psi_0}$ is the initial state which is assumed to lie within the subspace left invariant by $P$. The assumption of frequent measurements  implies that the response bandwidth of the measurement is very large and can be achieved by $N \gg 1$. This allows us to treat the natural evolution as a first order perturbation (in $\tau \equiv t/N$). Hence we can approximate the evolution by $\ket{\psi(t)} \approx e^{-iH_{eff}t}\ket{\psi_0}$, where $H_{eff}$ is an effective Hamiltonian: $H_{eff} = PHP$. More general and sophisticated derivations of the same result are in \cite{Misra_S_1977, Facchi_P_2001, Facchi_P_2002, Facchi_2002, Facchi_LP_2003}.  In the general case, the resulting system dynamics is a modified Hamiltonian evolution on a subspace with an irreversible component rapidly suppressing coherence between the subspace and its orthogonal complement. 

The point to note from the above is that we achieve an effective modified Hamiltonian dynamics for the system through its irreversible interaction with a measuring device with sufficiently fast response. An ideal ECWM procedure does exactly this. To see this, note that the general evolution of an encoded state coupled to an environment and undergoing ECWM is
\begin{widetext}
\begin{equation}
\ket{\phi(t)} = [(P_A \otimes I_S \otimes I_E)(U_{AS} \otimes I_E)(I_A \otimes e^{-iH_{SE}\frac{t}{N}})]^N \ket{0}_A\ket{\psi_0}_S\ket{e}_E
\label{eq:evol}
\end{equation}
\end{widetext}
where the subscripts A, S, and E stand for ancilla, system and environment, respectively. $\ket{\phi(t)}$ is the combined state of all three sub-systems. The initial state is assumed to be a product state of the three subsystems, and the initial system state, $\ket{\psi_0}$ is assumed to lie within the codespace, while the initial ancilla state is assumed to be a known fiducial state. The first operator in \erf{eq:evol} represents a coupling of the system to the environment - the error. We consider a completely general coupling, so 
\begin{equation}
H_{SE} = \sum_k A_k^{(S)} \otimes B_k^{(E)},
\end{equation}
and the operators $\{A_k\}$ are the errors on our system. The second operator in \erf{eq:evol} is a unitary operation between the system and ancilla subsystems which implements the error detection/correction, and the third is the ancilla reset operation which can be viewed as a projection of the ancilla onto their fiducial states - i.e. $P_A = \ket{0}_A\bra{0}$. We do not specify $U_{AS}$ or put restrictions  on the dimensions of the system and ancilla subspaces, except that they be finite, so this set-up could be implementing any error correction code. Note that the detect/correct and reset operations are assumed to be instantaneous while the error coupling is a Hamiltonian evolution. We will refer to the sequence within the square brackets in \erf{eq:evol} as a cycle. 

We are interested in the regime where the error correction operations are done frequently - when $N \gg 1$ and thus $\tau \equiv t/N \ll 1$. In this regime, the system-environment coupling is weak compared to the error correction operations and we can expand the exponential in the error operator to first order in $\tau$:
\begin{eqnarray}
\ket{\phi(t)} \approx [(P_A)(U_{AS})(I - i\tau H_{SE})]^N \ket{0}_A\ket{\psi_0}_S\ket{e}_E
\label{eq:pertexp}
\end{eqnarray}
Here we have suppressed the tensor product signs and dispensed with explicitly writing the identity operators. Also, for ease of notation let $PU \equiv P_AU_{AS}$ and $\ket{\phi_0} \equiv \ket{0}_A\ket{\psi_0}_S\ket{e}_E$. For the remaining derivation we will need the following property:

\textit{Property:} 
\begin{equation}
PUP \equiv P_AU_{AS}P_A = P_A\Pi_S,
\label{eq:prop}
\end{equation}
where $\Pi_S$ is projector onto the codespace in the encoded (system) subspace. 

\textit{Proof:}
First note that the subspace projected onto by $P_A\Pi_S$ is defined to be invariant (and furthermore, stabilized) by $U_{AS}$ - i.e. $[U_{AS}, P_A\Pi_S] = 0$, and $U_{AS}P_A\Pi_S = P_A\Pi_S$. Now letting letting $\Gamma_S = I_S - \Pi_S$, 
\begin{eqnarray}
P_AU_{AS}P_A &=& P_A(\Pi_S + \Gamma_S)U_{AS}P_A(\Pi_S + \Gamma_S) \nn \\
&=& P_A\Pi_S + P_A\Gamma_SU_{AS}P_A\Gamma_S
\end{eqnarray}
The second term on the last line above is a restriction of $U_{AS}$ to the subspace spanned by the projector $P_A\Gamma_S$: $\mathcal{H}_{P_A\Gamma_S}$. We will show that this is zero, and therefore prove the property. The fact that $P_A\Gamma_SU_{AS}P_A\Gamma_S=0$ follows from the definition of $U_{AS}$, which takes every vector in $\mathcal{H}_{P_A\Gamma_S}$ to a vector outside it. That is, if the encoded state is not in the codepace, $U_{AS}$ is defined to set the ancilla qubits to a state orthogonal to the fiducial state.

A corollary of property \erf{eq:prop} is that $(P_AU_{AS})^n = P_A\Pi_SU_{AS}$ for any integer $n>1$. Now, returning to \erf{eq:pertexp},
\begin{eqnarray}
\ket{\phi(t)} &=& [(PU)(I - i\tau H_{SE})]^N \ket{\phi_0} \nn \\
&\approx& (PU)^N\ket{\phi_0} - i\tau \sum_{k=1}^{N} (PU)^k H_{SE} (PU)^{N-k}\ket{\phi_0} \nn \\
&=& \ket{\phi_0} - i\tau \sum_{k=1}^{N} (PUP)^k H_{SE} (PUP)^{N-k} \ket{\phi_0} \nn \\
&=& \ket{\phi_0} - i\tau (N-1) (P_A\Pi_S) H_{SE} (P_A \Pi_S) \ket{\phi_0}
\label{eq:state}
\end{eqnarray}
where we have ignored all terms higher than first order in $\tau$. In the above we have used \erf{eq:prop}, $P^2 = P$, $P\ket{\phi_0} = \ket{\phi_0}$, $PU\ket{\phi_0} = \ket{\phi_0}$, and $[H_{SE},P_A] = 0$. Now, the second term in \erf{eq:state} is zero by design because:
\begin{eqnarray}
(P_A\Pi_S) H_{SE} (P_A \Pi_S) = P_A \otimes \sum_k \Pi_S A_k^{(S)} \Pi_S \otimes B_k^{(E)}
\end{eqnarray}
and the error correction code is designed so that $\Pi_S A_k^{(S)} \Pi_S=0$ for all $k$. This is a consequence of the error correction conditions/criteria \cite{Knill_L_1997, mikeandike}. Therefore, $\ket{\phi(t)} = \ket{\phi_0}$, and the encoded state is preserved. Note that just as in the Zeno effect, we can think of the system evolving according to the effective Hamiltonian $H_{eff} = (N-1) (P_A\Pi_S) H_{SE} (P_A \Pi_S) = 0$. And just as in the Zeno effect, this modified evolution depends strongly on the fact that error correcting operations occur frequently and are much stronger than the interaction/error Hamiltonian, $H_{SE}$. 


In closing, we note that this connection between ECWM with ideal resources and the Zeno effect has been used in \cite{Vaidman_GW_1996, Erez_ARV_2003} to construct error \textit{prevention} techniques that use fewer resources than error correction codes.

%
%


\bibliography{ecpzv7}

\end{document}